\documentstyle[prl,aps,floats,psfig]{revtex}
\psfigurepath{/homes/solid/radtke/supercon/bcscalcs}

\begin{document}
\twocolumn[
\hsize\textwidth\columnwidth\hsize\csname@twocolumnfalse\endcsname

\title{Antiferromagnetic Interactions
and the Superconducting Gap Function}
\author{R. J. Radtke,$^1$ A. I. Liechtenstein,$^2$
V. M. Yakovenko,$^1$ and S. Das Sarma$^1$}
\address{$^1$Center for Superconductivity Research, Department of Physics,
University of Maryland, College Park, Maryland  20742-4111\\
$^2$Max-Planck-Institut f\"{u}r Festk\"{o}rperforschung,
Heisenbergstrasse 1, D-70569 Stuttgart, Federal Republic of Germany}
\date{Published as Phys. Rev. B {\bf 53}, 5137 (1996) [cond-mat/9507013]}
\maketitle

\begin{abstract}
Spin-fluctuation-mediated superconductivity is conventionally
associated with $d_{x^2-y^2}$ pairing.  We show that a
generalized model of antiferromagnetic spin fluctuations in
three dimensions may also yield a state with formal ``$s$-wave''
($A_{1g}$) symmetry but with line nodes at $k_z \approx \pm \pi / 2c$.
We study this new state within both BCS and Eliashberg theories
using a realistic band structure and find that it is more stable
than the $d_{x^2-y^2}$ ($B_{1g}$) state over a wide range of
parameters.  Thus, models of spin-fluctuation-mediated
superconductivity must consider both possibilities on an equal footing.\\
\\
PACS numbers:  74.20.-z, 74.72.-h, 74.25.Dw, 74.20.Fg\\
\\
\end{abstract}

]

One of the fundamental questions concerning the cuprate superconductors
is the symmetry of the superconducting gap function $\Delta_{\bf k}$.
Experiments which probe the relative sign of the Josephson
currents in the $a$ and $b$ directions in $\rm YBa_2Cu_3O_7$
(YBCO) \cite{dSQUID} are commonly considered proof of $d_{x^2-y^2}$ pairing.
However, recent theories which account for the CuO chains in YBCO
have challenged this interpretation \cite{Mazin}, and 
other measurements exist \cite{sSQUID} which are difficult to reconcile
with this pairing symmetry.
Thus, current experiments do not unambiguously establish
$d_{x^2-y^2}$ pairing for all the cuprates under all conditions.

The experiments of Ref.~\onlinecite{dSQUID} are consistent with
theories of high-temperature superconductivity based on
antiferromagnetic spin fluctuations \cite{Scalapino,Ueda,Pines}.
Since the CuO$_2$ planes in the cuprates are generally considered
the most important structural elements for the superconductivity,
most work on this subject has focussed on a single, two-dimensional (2D)
band, representing the physics of an isolated CuO$_2$ plane.
Within this framework, it is now well established that antiferromagnetic
spin fluctuations yield singlet pairing in a $d_{x^2-y^2}$-like
orbital state \cite{Scalapino,Ueda,Pines}.
This pairing symmetry can be understood by noting that
antiferromagnetic spin fluctuations produce a repulsive interaction
which inhibits on-site pairing but favors pairing on
nearest-neighbor sites \cite{Scalapino2}.

In materials like YBCO, though, there are {\it two} CuO$_2$ planes
per unit cell.
One therefore expects that {\it two} 2D bands will be involved
in the physical description, both of which may contribute to the
superconductivity.
Recent calculations within such bilayer systems based on either
weakly \cite{Bulut,Liechtenstein,Liu} or strongly \cite{Ioffe,Lee}
correlated electrons concluded that an alternative
to the $d_{x^2-y^2}$ ($B_{1g}$) state may exist.
In this state, there are two gap functions (one for
each band) which may be strongly anisotropic in wave vector space
but which do not have nodes; however, the two gap functions do
have opposite signs.
Hence, this state has been termed the $s^{\pm}$ state \cite{MY}.
Strong-coupling Eliashberg calculations have demonstrated that
the $s^{\pm}$ state is actually more stable than the $d_{x^2-y^2}$
state whenever the inter-band antiferromagnetic correlations are
stronger than the intra-band ones \cite{Liechtenstein}.
This pairing symmetry may also be consistent with all
\cite{dSQUID,sSQUID} of the experimental determinations of the
phase of $\Delta_{\bf k}$ \cite{Mazin}.

The $s^{\pm}$ state is restricted to the bilayer cuprates such as
YBCO, and thus does not address the superconductivity in single-layer
cuprates like $\rm La_{1-x}Sr_xCuO_4$ (LSCO).
Moreover, both $s^{\pm}$ and $d_{x^2-y^2}$ states are based on a 2D
picture that neglects the third dimension.
In particular, increasing the bilayer separation
in a three-dimensional (3D) layered system causes the two-band
structure to cross-over into a single-band one.
Thus, there should be a one-band analog of the $s^{\pm}$ state
in 3D which would compete with the $d_{x^2-y^2}$ state.

We examine this possibility within a general model
for antiferromagnetic spin fluctuations in a single 3D band.
By solving both the weak-coupling BCS and strong-coupling
Eliashberg gap equations at $T_c$, we are able to compute
the phase diagram for the different pairing symmetries.
We find that the one-band analog of the $s^{\pm}$ state exists and
is stable over a large range of antiferromagnetic correlation
lengths which parameterize our model.
We emphasize that, even if fully oxygenated YBCO has $d_{x^2-y^2}$
pairing, one may be able to observe a transition to this $s^{\pm}$
state by changing the correlation lengths through changes in
oxygen content, pressure, and chemical substitution.
The possible experimental relevance of this state therefore
demands that it be considered as a serious alternative to the
conventional $d_{x^2-y^2}$ state in future calculations.

To be as general as possible, we take the antiferromagnetic
spin fluctuation interaction to have the form
\begin{equation}
V ({\bf q}, \omega) = \frac{V_0}
  {1 + \xi_{ab}^2 ({\bf q}_{\parallel} - {\bf Q}_{\parallel})^2
     + \xi_{c}^2 (q_{\bot} - Q_{\bot})^2 - i\omega/\omega_{\rm SF}},
\label{eq:V}
\end{equation}
where $V_0$ is the electron-spin fluctuation coupling strength;
$({\bf Q_{\parallel}}, Q_{\bot}) = (\pi/a,\pi/a,\pi/c)$
and $\omega_{\rm SF}$ = 7.7 meV are the spin fluctuation wave vector
and characteristic frequency; and  $\xi_{ab}$ and $\xi_{c}$ are the
intra- and inter-planar  antiferromagnetic correlation lengths, respectively.
[Throughout this paper, $a$ ($c$) is the unit cell dimension normal
to (along) the $z$ direction.]
When $\xi_c \rightarrow$ 0, Eq.~(\ref{eq:V}) reduces to the 
2D spin fluctuation model  of Millis, Monien, and Pines \cite{MMP,Pines}.
$\xi_c$ is introduced into this model in a straightforward way
to account for the experimental fact that inter-planar
antiferromagnetic correlations exist in YBCO \cite{neutron}.

This interaction [Eq.~(\ref{eq:V})] is assumed to exist between
quasiparticles which have a 3D dispersion relation $\epsilon_{\bf k}$.
To further relate our calculations to the cuprates, we take
$\epsilon_{\bf k}$ from the single-band model developed
by Andersen {\it et al.} for YBCO \cite{Andersen}:
\begin{eqnarray}
\epsilon_{\bf k} &=& -2 t_1 ( \cos k_xa + \cos k_ya)
  - 4t_2 \cos k_xa \cos k_ya \nonumber \\
  &&-\frac{t_3}{2} (\cos k_xa - \cos k_ya)^2 \cos k_zc - \mu.
\label{eq:bs}
\end{eqnarray}
Here, $t_1$ = 0.25 eV,  $t_2$ = -0.1 eV, and $t_3$ = 0.05 eV are
the matrix elements for intra-planar nearest-neighbor,
intra-planar next-nearest-neighbor, and inter-planar hopping,
respectively; $\mu$ = -0.4 eV is the chemical potential.
Note the surprising dependence of the inter-planar hopping on the
intra-planar wave vector.
This behavior results from the reduction of the full local-density
approximation band structure to a single-band, tight-binding
dispersion; see Ref. \onlinecite{Andersen} for details.
Equation~(\ref{eq:bs}) reproduces the general features of the
band structure of YBCO revealed by angle-resolved photoemission
experiments \cite{ARPES} and yields the Fermi surface shown
in Fig.~\ref{fig:gapfns}.

\begin{figure}[t]
\parbox{\linewidth}{
\rule{0in}{1.1\linewidth}
}
\caption{Gap functions $\Delta_{\bf k}$ for a three-dimensional
antiferromagnetic spin-fluctuation-mediated superconductor
[Eq.~(\protect\ref{eq:V})] which transform under the
(a) $B_{1g}$ $(d_{x^2-y^2}$) and (b) $A_{1g}$ ($s^{\pm}$)
representations of the tetragonal point group.
The Fermi surface in the first Brillouin zone [Eq.~(\protect\ref{eq:bs})]
is shown colored by the value of $\Delta_{\bf k}$ on the Fermi
surface, with blue denoting positive values and red denoting
negative ones.
Observe the different locations of the gap nodes (white stripes) around
the Fermi surface.}
\label{fig:gapfns}
\end{figure}

With the pairing interaction and the band structure specified, we can proceed
to study the superconductivity within either BCS theory or its
strong-coupling counterpart, Eliashberg theory \cite{Eliashberg}.
For simplicity, let us first consider the BCS equation at $T_c$,
\begin{equation}
\lambda_{\rm pair} \Delta_{\bf k} =
  - \frac{1}{N} \sum_{\bf k'} V_{{\bf k} - {\bf k'}} \,
  \frac{{\rm tanh} (\epsilon_{\bf k'} / 2k_BT)}{2\epsilon_{\bf k'}} \,
  \Delta_{\bf k'},
\label{eq:bcs}
\end{equation}
where $\Delta_{\bf k}$ is the gap function,
$V_{\bf q} = V ({\bf q}, 0)$ is the pairing interaction [Eq.~(\ref{eq:V})],
$\epsilon_{\bf k}$ is the electronic dispersion [Eq.~(\ref{eq:bs})],
$N$ is the number of sites in the lattice,
$T$ is the temperature,
and $\lambda_{\rm pair}$ is the pairing eigenvalue.
This equation is written for singlet pairing, for which
$\Delta_{-{\bf k}} = \Delta_{\bf k}$; for triplet pairing,
$\Delta_{-{\bf k}} = -\Delta_{\bf k}$, and the right-hand side of
Eq.~(\ref{eq:bcs}) is multiplied by -1/3.
The Eliashberg equations are similar in spirit although formally
more complicated; the interested reader is referred to the
literature \cite{Eliashberg}.

Equation~(\ref{eq:bcs}) is an eigenvalue equation, allowing the
eigenvectors $\Delta_{\bf k}$ to be classified into different
representations of the crystal point group by their transformation
properties under the elements of that group \cite{Hamermesh,SU}.
The ``pairing symmetry'' of $\Delta_{\bf k}$ refers to this representation.
For example, in tetragonal systems (point group $D_{4h}$),
``$d_{x^2-y^2}$ pairing'' corresponds to the $B_{1g}$ representation and
``$s^{\pm}$ pairing'' to the $A_{1g}$ representation.
To determine the most favored pairing symmetry, one can either
find the representation yielding the largest $\lambda_{\rm pair}$
at fixed $T$ or the largest $T_c$ (defined by
$\lambda_{\rm pair} (T = T_c) = 1$).
We employ both methods to construct the phase diagram for our model
and find consistent results.

We solve Eq.~(\ref{eq:bcs}) and its strong-coupling counterpart using
a fast Fourier transform technique with the wave vector integrations
carried out over the entire Brillouin zone \cite{FZ}.
The extremal eigenvalues for each one-dimensional representation
of the point group are obtained by an iterative procedure with
an eigenvalue shift \cite{Acton}.
The BCS calculations are carried out on a 32 x 32 x 32 discretization
of the Brillouin zone, and the strong-coupling calculations are
performed on a 64 x 64 x 8 mesh with 256 Matsubara frequencies,
corresponding to a frequency cut-off on the order of three times
the band width.
Except as noted below, our results are insensitive to these choices.

In discussing our results, we begin with an examination of the
gap function symmetries associated with the dominant pairing eigenvalues.
We have calculated $\lambda_{\rm pair}$ for all the one-dimensional
representations of the tetragonal point group over a large range of correlation
lengths (see below), and we find only two representations which lead to
a superconducting instability:  $B_{1g}$ and $A_{1g}$.
The $B_{1g}$ solution is shown in Fig.~\ref{fig:gapfns}(a) and corresponds
to the $d_{x^2-y^2}$-like solution encountered in previous work
on a single band in 2D \cite{Scalapino,Ueda,Pines}.
This state possesses nodal planes defined by the directions [110] and [001]
and by [1$\overline{1}$0] and [001],
and can be thought of as an intra-planar pairing of quasiparticles on
nearest-neighbor sites \cite{Scalapino2}.
On the other hand, the $A_{1g}$ solution is new.
From Fig.~\ref{fig:gapfns}(b), we see that the $A_{1g}$ solution, while
formally having ``$s$-wave'' (i.e., $A_{1g}$) symmetry, possesses
nodal lines at $k_z \approx \pm \pi / 2c$ \cite{note}.
The gap function $\Delta_{\bf k}$ thus transforms like $\Delta_{\bf k} \sim \cos k_zc$
and corresponds to the pairing of quasiparticles between {\it different planes}.
This solution is the one-band analog of the $s^{\pm}$ state discussed
within the context of two-band, 2D models
\cite{Bulut,Liechtenstein,Liu,Ioffe,Lee}.
We can see this analogy by mentally translating every other plane
in this monolayer model to produce a bilayer; under this transformation,
the regions of $\Delta_{\bf k}$ with the same sign in Fig.~\ref{fig:gapfns}(b)
would be mapped onto one of the resulting two bands, and the nodes
would disappear in favor of the band gap, maximizing the superconducting
condensation energy.

The one-band $s^{\pm}$ state exists as a possible
superconducting solution, but is it stable?
When $\xi_c \rightarrow$ 0, Eq.~(\ref{eq:bcs}) indicates that
$\Delta_{\bf k}$ can only depend on $k_x$ and $k_y$;  we therefore
expect that the only stable solution will be the $d_{x^2-y^2}$-like
($B_{1g}$) one.
On the other hand, when $\xi_{ab} \rightarrow 0$, Eq.~(\ref{eq:bcs})
must yield a $\Delta_{\bf k}$ which depends only on $k_z$.
Moreover, we require a sign change in the $\Delta_{\bf k}$ connected by
wave vectors where $V_{\bf q}$ is maximal (at ${\bf q} = (0,0,\pi/c)$)
in order to cancel the negative sign on the right-hand side of Eq.~(\ref{eq:bcs}).
These demands are met by the $s^{\pm}$ state of Fig.~\ref{fig:gapfns}(b),
and so this solution should be stable in this limit.
For intermediate values of the correlation lengths, we should see
a cross-over between these states \cite{note00}.

\begin{figure}[tb]
\centerline{
\psfig{figure=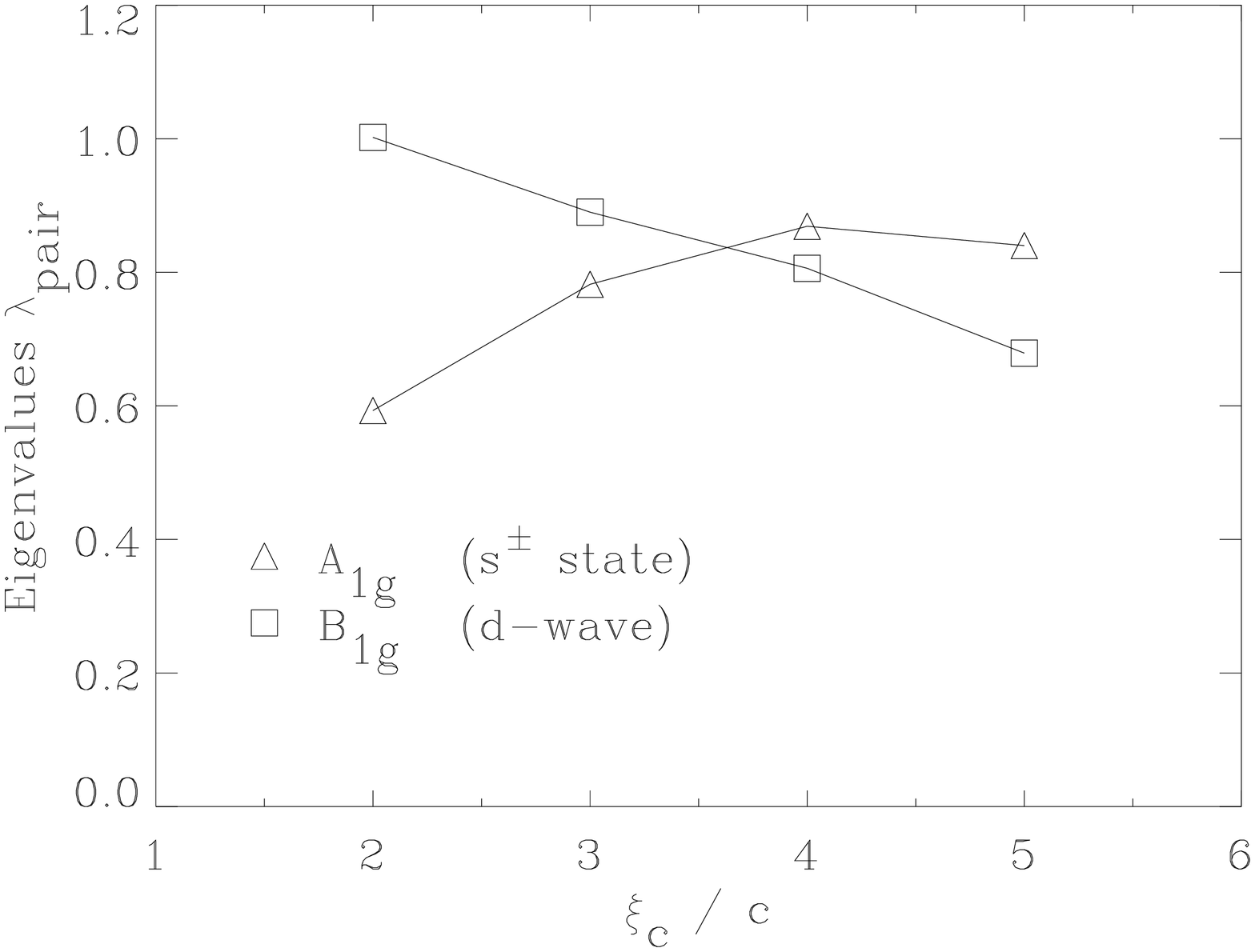,width=0.95\linewidth,angle=90}}
\caption{Pairing eigenvalues $\lambda_{\rm pair}$ belonging to
the $B_{1g}$ $(d_{x^2-y^2}$, squares) and $A_{1g}$
($s^{\pm}$, triangles) representations as a function
of the normalized inter-planar antiferromagnetic correlation
length $\xi_c / c$; the solid lines are to guide the eye.
These results are computed within strong-coupling Eliashberg theory
for an antiferromagnetic spin-fluctuation-mediated interaction
[Eq.~(\protect\ref{eq:V})] with an intra-planar correlation length
$\xi_{ab}$ = 2.3$a$ at 90~K.
Note the crossover between $B_{1g}$ and $A_{1g}$ pairing around
$\xi_c$ = 3.5$c$, which is consistent with the BCS result of
Fig.~\protect\ref{fig:bcstc}.}
\label{fig:sceval}
\end{figure}

\begin{figure}[tb]
\centerline{
\psfig{figure=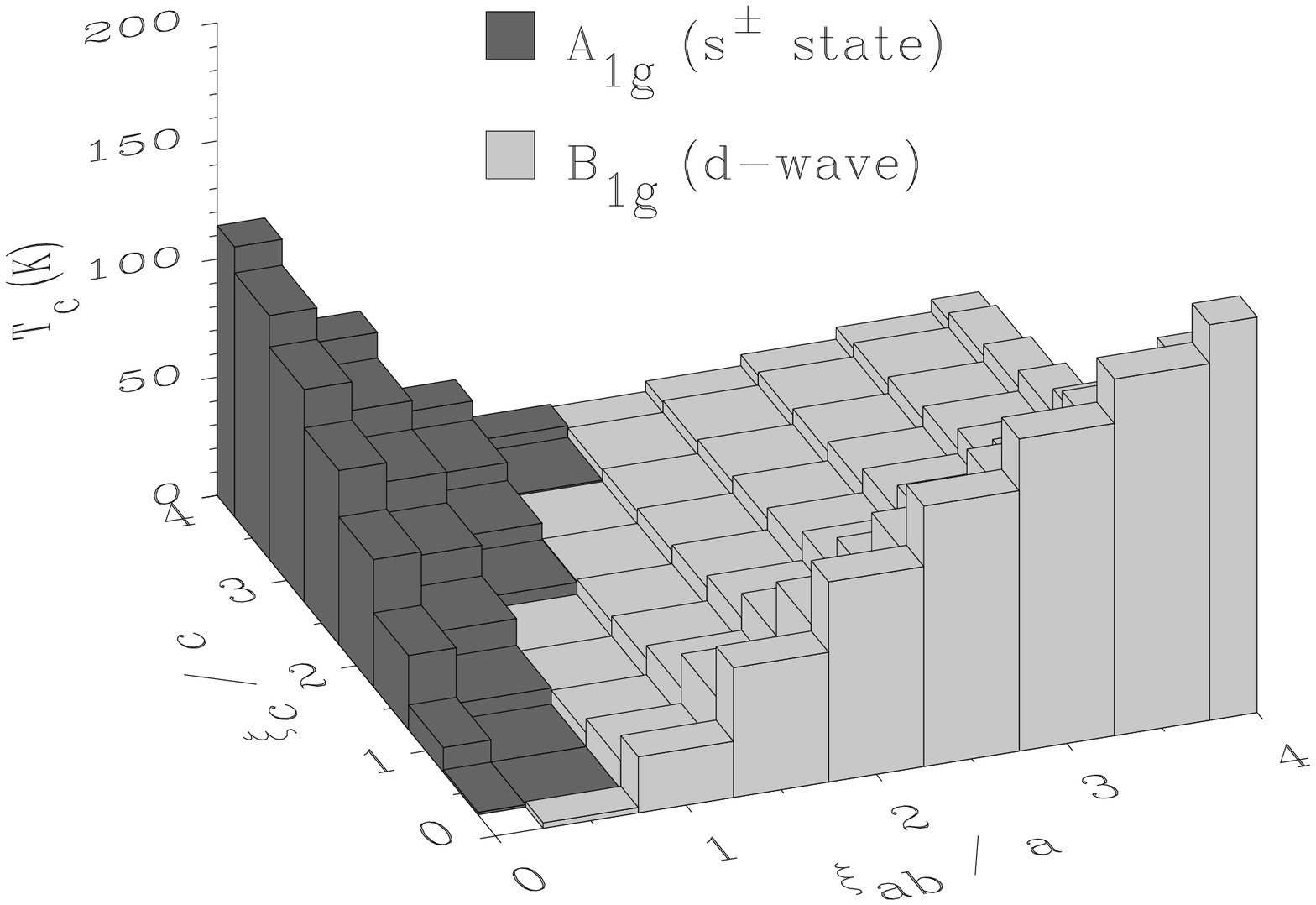,width=0.95\linewidth,angle=90}}
\caption{BCS critical temperature $T_c$ of
an antiferromagnetic spin-fluctuation-mediated superconductor
[Eqs.~(\protect\ref{eq:V})-(\protect\ref{eq:bcs})] as a function of the normalized
inter- ($\xi_c / c$) and intra-planar ($\xi_{ab} / a$)
correlation lengths.
Light shading represents $T_c$ for
$B_{1g}$ $(d_{x^2-y^2}$) pairing while dark shading
represents $A_{1g}$ ($s^{\pm}$) pairing.}
\label{fig:bcstc}
\end{figure}

We have tested these expectations numerically and present our results
in Figs.~\ref{fig:sceval} and \ref{fig:bcstc}.
Consider first Fig.~\ref{fig:sceval}, which shows $\lambda_{\rm pair}$
as a function of $\xi_c$ for $\xi_{ab}$ = 2.3a computed within
strong-coupling Eliashberg theory.
The largest pairing eigenvalues, and hence the most favored
superconducting solutions, belong to the
$B_{1g}$ ($d_{x^2-y^2}$) or $A_{1g}$ ($s^{\pm}$) representations.
As $\xi_c$ increases, we see that the most stable solution
crosses over from the $B_{1g}$ to the $A_{1g}$ representation, as
we expect from the preceding discussion.

To give these results a more physical interpretation, we show in
Fig.~\ref{fig:bcstc} the critical temperature $T_c$ of the most stable
pairing symmetry as a function of the intra- and inter-planar correlation
lengths determined from a BCS calculation \cite{note2,note3}.
The  $A_{1g}$ state is stable over a large region of
parameter space, with the phase boundary given roughly
by the line $\xi_c / c$ = 2 $\xi_{ab} / a$.
The limiting cases discussed above are confirmed as well:
the $B_{1g}$ ($d_{x^2-y^2}$) solution dominates as $\xi_c \rightarrow 0$,
and the reverse is true for the $A_{1g}$ ($s^{\pm}$) state as
$\xi_{ab} \rightarrow 0$.
In addition, the critical temperatures are largest along the $\xi_c = 0$
and $\xi_{ab} = 0$ lines and increase as the antiferromagnetic
correlation lengths increase.

Moving towards the phase boundary in Fig.~\ref{fig:bcstc},
$T_c$ decreases but does not vanish, at least for large $\xi_{ab}$ and $\xi_c$.
At small correlation lengths, however, $T_c$ is small near the phase
boundary and very sensitive to the finite-size effects
induced by our discretization of the Brillouin zone.
Hence, we cannot exclude the possibility of a finger of normal
phase extending between the two superconducting phases at
small $\xi_{ab}$ and $\xi_c$.
With this caveat, we believe the Fig.~\ref{fig:bcstc} represents
the true phase diagram of our model.

It is important to note that the existence of the one-band
$s^{\pm}$ ($A_{1g}$) state depends only on the 3D
interaction and not the detailed band structure.
Specifically, recomputing the phase diagram with $t_3 = 0$ in
Eq.~(\ref{eq:bs}) gives nearly identical results.
The $s^{\pm}$ state may therefore be compatible with theories of the
cuprate superconductors in which single-particle inter-layer
hopping is suppressed but Cooper pair tunneling is
not \cite{Anderson}.
Of course, the stability of the $s^{\pm}$ state in any concrete system,
as opposed to its existence, depends on microscopic details
such as the band structure and must therefore be evaluated on
a case-by-case basis.

To conclude, we have shown that single-band models of
antiferromagnetic spin fluctuation-mediated superconductivity
in 3D do {\it not} automatically give rise to $d_{x^2-y^2}$ pairing.
Rather, a phase which possesses formal ``$s$-wave'' ($A_{1g}$) symmetry
with nodal lines at $k_z \approx \pm \pi / 2c$ exists and is
stable over a wide range of parameters.
Within the context of the cuprate superconductors, these
results imply that this unusual $s$ state, which is the one-band analog
of the $s^{\pm}$ state of Ref.~\onlinecite{Liechtenstein}, must
be considered on equal footing with the conventional $d_{x^2-y^2}$ state.
This point is reinforced by the fact that the $s^{\pm}$ state is favored
when the inter-planar antiferromagnetic correlations are stronger than
the intra-planar ones, a situation which may obtain in the cuprates.
Finally, the results of this paper suggest that, even in a superconductor
with pure $d_{x^2-y^2}$ pairing, substitution, doping, or
pressure may lead to a change in the correlations lengths
and thus to a transition between different gap function symmetries.
Such transitions may already have been observed in
angle-resolved photoemission experiments \cite{Onellion}.

We would like to thank I. I. Mazin for helpful discussions.
This work was supported in part by the NSF under Grants
DMR-9123577 (RJR, SDS) and DMR-9417451 (VMY) and by the
A.~P.~Sloan Foundation (VMY).


\begin{references}
\bibitem{note}Numerically, the nodes are located at $k_z = \pm \pi / 2c$
for the gap functions we have examined, but there is no group theoretic
reason that they {\it must} be at those locations.
\bibitem{note00}The point ($\xi_{ab}$,$\xi_c$) = (0,0) is exceptional
in that there can be no solution for any representation when $V_0 > 0$
[cf. Eq.~(\ref{eq:bcs})].
\bibitem{note2}
$T_c$ is easier to compute in BCS theory than in
Eliashberg theory, but is less accurate.
Nonetheless, the phase boundary from the BCS calculations is
in close agreement with the Eliashberg results of Fig.~\protect\ref{fig:sceval},
which supports the qualitative correctness of the BCS phase diagram.
\end{references}
\end{document}